\documentclass[prd,showpacs,superscriptaddress,dblfloatfix,preprintnumber,floatfix,footinbib]{revtex4-2}

\usepackage{graphicx} % Required for inserting images
\pdfoutput=1

\usepackage{amsmath,amssymb}
\usepackage{epsfig}
\usepackage{hyperref}
\usepackage[dvipsnames]{xcolor}
\usepackage{slashed}
\usepackage{placeins}

\usepackage{amsfonts}
\usepackage{graphicx, rotating}
\usepackage{epstopdf}
\usepackage{latexsym}
\usepackage{rotating}
\usepackage{braket}
\usepackage{multirow}
\usepackage{makecell}

\usepackage{subcaption}
\usepackage[export]{adjustbox}
\usepackage[version=3]{mhchem}
\usepackage{makecell}
\usepackage{xcolor}

\usepackage[font={small}]{caption}   % Writes captions of figures in small font
\usepackage{lineno}

\begin{document}
%\linenumbers

\title{Searching for Ultralight Dark Matter with M\"ossbauer Resonance}

\author{Peng-Long Zhang}
%\email{plzhang@ihep.ac.cn}
\affiliation{Centre for Theoretical Physics, Henan Normal University, Xinxiang 453007, China}
\affiliation{Institute of High Energy Physics, Chinese Academy of Sciences, Beijing 100049, China}

\author{Yu-Ming Yang}
\affiliation{Institute of High Energy Physics, Chinese Academy of Sciences, Beijing 100049, China}
\affiliation{School of Physical Sciences, University of Chinese Academy of Sciences, Beijing 100049, China}

\author{Xiao-Jun Bi}
%\email{bixj@ihep.ac.cn}
%\affiliation{Key Laboratory of Particle Astrophysics, Institute of High Energy Physics, Chinese Academy of Sciences, Beijing 100049, China}
\affiliation{Institute of High Energy Physics, Chinese Academy of Sciences, Beijing 100049, China}
\affiliation{School of Physical Sciences, University of Chinese Academy of Sciences, Beijing 100049, China}

\author{Qin Chang}
%\email{changqin@htu.edu.cn}
\affiliation{Centre for Theoretical Physics,
Henan Normal University, Xinxiang 453007, China}
\affiliation{Center for High Energy Physics, Henan Academy of Sciences, Zhengzhou 455004, China}

\author{Yu Gao}
\affiliation{Institute of High Energy Physics, Chinese Academy of Sciences, Beijing 100049, China}

\author{Hai-Bo Li}
\affiliation{Institute of High Energy Physics, Chinese Academy of Sciences, Beijing 100049, China}
\affiliation{School of Physical Sciences, University of Chinese Academy of Sciences, Beijing 100049, China}

%\author{Xing-Jian Lv$^{\P}$}
%\affiliation{Institute of High Energy Physics, Chinese Academy of Sciences, Beijing 100049, China}
%\affiliation{School of Physical Sciences, University of Chinese Academy of Sciences, Beijing 100049, China}

\author{Wei Xu}
\affiliation{Institute of High Energy Physics, Chinese Academy of Sciences, Beijing 100049, China}

\author{Peng-Fei Yin}
%\email{yinpf@ihep.ac.cn}
\affiliation{Institute of High Energy Physics, Chinese Academy of Sciences, Beijing 100049, China}

\date{\today}

\begin{abstract}
We investigate the feasibility of probing the interactions between ultralight scalar dark matter and atomic nuclei using a stationary Mössbauer spectroscopy scheme. The exceptional energy resolution of the Mössbauer resonance enables searches for tiny nuclear energy shifts induced by the local dark matter field. 
The dark matter mass range considered in this work is $10^{-18}$--$10^{-8}~\mathrm{eV}$. We present projected constraints for two candidate Mössbauer isotopes, $^{109}\mathrm{Ag}$ and $^{45}\mathrm{Sc}$, with $^{109}\mathrm{Ag}$ providing the strongest sensitivity. For $^{109}\mathrm{Ag}$, projected sensitivities as low as approximately $10^{-19}$, $10^{-22}$, and $10^{-21}~\mathrm{GeV^{-1}}$ can be achieved for the scalar DM--photon, DM--gluon, and DM--quark couplings $f_{\gamma}^{-1}$, $f_{g}^{-1}$, and $f_{\hat{m}}^{-1}$, respectively.
In the low-mass region, the projected sensitivity to the scalar DM--photon coupling approaches the current constraints from equivalence-principle (EP) tests.
These results demonstrate that Mössbauer-based techniques provide a promising and competitive approach for probing ultralight dark matter interactions with Standard Model particles.

%We present the projected sensitivity to the dark matter parameter space for two candidate Mössbauer isotopes, $^{109}\text{Ag}$ and $^{45}\text{Sc}$. Among them, $^{109}\text{Ag}$ provides the highest sensitivity, while $^{45}\text{Sc}$ yields comparatively weaker constraints. For $^{109}\text{Ag}$, the scalar dark-matter–photon coupling $f_{\gamma}^{-1}$ can be constrained down to the level of $10^{-21}~\mathrm{GeV^{-1}}$, exceeding the sensitivity of equivalence principle(EP) experiments. The scalar dark-matter–gluon coupling $f_{g}^{-1}$ can be probed down to $10^{-24}~\mathrm{GeV^{-1}}$ and exceeding the EP test constrain, while the scalar dark-matter–quark coupling $f_{\hat{m}}^{-1}$ can reach approximately $10^{-23}~\mathrm{GeV^{-1}}$. These results demonstrate that Mössbauer-based techniques offer a promising and competitive approach for probing ultralight dark matter interactions with Standard Model particles. 

%In principle, a stationary measurement allows faster data acquisition and becomes advantageous at higher dark-matter masses in the range $10^{-16}$–$10^{-8}~\mathrm{eV}$.

%\ck{\bf [give your main result quantitatively], and add a brief comment on the most attractive point of your result.}
 
%This study highlights the potential of Mössbauer spectroscopy as a sensitive probe for new scalar interactions at submicrometer scales and demonstrates its feasibility as a novel method for dark matter detection.

\end{abstract}

\maketitle

\section{Introduction}
Understanding the fundamental nature of dark matter remains one of the most profound open questions in modern physics. Although dark matter constitutes approximately 26.8\% of the total energy density of the Universe--making it roughly five times more abundant than ordinary (baryonic) matter~\cite{Planck:2018vyg}--its microscopic properties and interactions continue to evade detection. The evidence for dark matter primarily arises from its gravitational effects on galactic and cosmological scales~\cite{Planck:2018vyg, Corbelli:1999af, Clowe:2006eq, Clowe:2003tk, Markevitch:2001ri, Jungman:1995df}, yet all attempts to directly observe its non-gravitational interactions with Standard Model particles have so far yielded null results.

Among the wide landscape of dark matter candidates, ultralight dark matter (ULDM) has attracted increasing attention. This class of models postulates bosonic fields with masses spanning a wide range, from $m \sim 10^{-23}\rm{eV}$ up to $1~\rm{eV}$~\cite{Ferreira:2020fam}. In many realizations--such as axions, axion-like particles, and scalar dark matter--the ULDM field behaves as a coherently oscillating classical wave, giving rise to distinctive experimental signatures. These particles often couple feebly to photons, electrons, or nucleons, potentially inducing time-varying fundamental constants or energy-level shifts in atomic and nuclear systems~\cite{Safronova:2017xyt,Antypas:2019yvv,Damour:1994zq,Barrow:1999is,Sandvik:2001rv,Arvanitaki:2014faa,Stadnik:2015kia,Banerjee:2018xmn,Hees:2018fpg,Savalle:2019jsb}. %\ck{\bf{add citations here}}.

The Mössbauer effect~\cite{Mossbauer:1958wsu} is well known for its exceptional energy resolution in recoil-free nuclear gamma-ray resonance. A classic application is the experimental verification of the gravitational redshift of photons in the Earth's gravitational field~\cite{KATILA198151, Potzel1992}. Over the years, the Mössbauer effect has had a profound impact in materials science and chemistry.
Its extremely high energy resolution also makes it a promising tool for probing tiny shifts in nuclear transition energies induced by couplings to a time-varying dark matter (DM) field. Such couplings can lead to small but measurable modifications of nuclear binding energies, resulting in shifts of the resonance lines in the Mössbauer spectrum. Importantly, the magnitude of these shifts is determined by the strength of the new interaction, while electromagnetic effects on nuclear energy levels are significantly suppressed due to shielding by the electron cloud. In addition, electromagnetic contributions typically couple to nuclear moments, which are themselves suppressed by the small size of the nucleus.

Recently, Refs.~\cite{Gratta:2020hyp,Gao:2023ggo,Banerjee:2024bkp} have explored the use of the Mössbauer effect as a tool to search for new physics. In Ref.~\cite{Gratta:2020hyp}, the authors propose exploiting the ultra-high energy resolution of Mössbauer spectroscopy to probe short-range interactions via tiny shifts in nuclear energy levels. They consider two complementary implementations: traditional radioactive Mössbauer sources and synchrotron-radiation-based excitation schemes. This approach significantly enhances sensitivity to new interactions at nanometer to micrometer scales and provides a unique probe of the underlying microscopic couplings and nuclear structure effects.
Ref.~\cite{Banerjee:2024bkp} focuses on using synchrotron-radiation Mössbauer spectroscopy to detect ultra-light scalar dark matter through modulations in the absorption spectrum over large source–absorber baselines.
Ref.~\cite{Gao:2023ggo} proposes a stationary Mössbauer scheme based on the $^{109}\mathrm{Ag}$ isotope. In this setup, perturbations from high-frequency gravitational waves under the local gravitational field can induce vertical shifts in the Mössbauer resonance height. The authors demonstrate that the ultra-high energy resolution of Mössbauer spectroscopy enables competitive sensitivity to gravitational waves across a wide frequency range, from KHz to above MHz.
In this work, we explore the sensitivity of ultralight scalar dark matter searches using a recently proposed static scheme~\cite{Gao:2023ggo}. Our analysis considers two radioactive Mössbauer isotopes, $^{109}\mathrm{Ag}$ and $^{45}\mathrm{Sc}$, as well as an X-ray free-electron-laser (XFEL)-based Mössbauer source using $^{45}\mathrm{Sc}$.

The rest of this paper is organized as follows. In Sec.~\ref{sec:DM_energy_shift_nucleus}, we describe the dark matter induced shift in nuclear energy levels and discuss the choice of Mössbauer nuclei used in our analysis. In Sec.~\ref{sec:Stationary_Scheme}, we introduce the stationary Mössbauer spectroscopy scheme for dark matter searches. In Sec.~\ref{sec:DM_signal}, we use Monte Carlo simulations and a $\chi^2$ analysis to estimate the dark matter induced shift in the absorber's resonance position, represented by a vertical displacement $\Delta Z_0$. In Sec.~\ref{sec:DM_Nuc_interact}, we present the interaction Lagrangian for dark matter couplings to photons, gluons, and quarks, and calculate the resulting small shifts in nuclear energy levels. In Sec.~\ref{sec:Projected_Senitivity}, we combine theoretical estimates with Monte Carlo simulations to derive projected constraints on the coupling parameters between dark matter and photons, gluons, and quarks. Finally, in Sec.~\ref{sec:conclusion}, we present our conclusions.

\section{Dark matter induced energy shifts and nuclear selection}
\label{sec:DM_energy_shift_nucleus}
In the ultralight bosonic dark matter scenario, the galactic dark matter behaves as a coherently oscillating classical scalar field with a central frequency set by the nonrelativistic particle energy $\omega_\phi=m_{\phi}+{\cal O}(\beta^2)$ and correspondingly a quite small energy dispersion $\sigma(\omega)\sim \beta^2 m_{\phi}$. Within one coherent time scale, the local dark matter field value can be written as
\begin{eqnarray}\label{eq:dark_matter01}
\phi(t,\vec x)\simeq\frac{\sqrt{2 \rho_{\rm DM}}}{m_\phi}\cos\left(\omega_\phi t-m_\phi\vec\beta\cdot\vec x\right)\,,
\end{eqnarray}
where $\rho_{\text{DM}} \sim 0.4\, \text{GeV/(cm)}^3$ is the local dark matter density and $|\beta|\sim 10^{-3}$ is the virial velocity, and $\omega_\phi\approx m_\phi$.
The shift of the nuclear transition energy induced by dark matter is proportional to the dark matter field. Thus, the energy-level shift of the emitter at position $\vec x=\vec 0$ at time $t$ is $\kappa \phi(t, \vec 0)$. When the photon reaches the absorber located at 
$\vec x=\vec L$, the corresponding shift is $\kappa \phi(t + L, \vec L)$.
% If a nuclear transition energy $E_{\gamma}$ depends linearly on the field ($E_{\gamma} = \kappa \phi(t,\vec x)$), then a photon emitted at time $t$ has energy $\kappa \phi(t, \vec 0)$. Upon arrival at an absorber located a distance $L$ away, the transition energy at the absorber is $E_{\gamma} = \kappa \phi(t + L, \vec L)$. 
The resulting difference between the emitter and absorber transition energies is given by
\begin{eqnarray}\label{eq:Energy_Shift}
\Delta E 
&=& \kappa \phi(t + L, \vec L) - \kappa \phi(t, \vec 0) \\ \nonumber
&=& \kappa \frac{\sqrt{2 \rho_{\rm DM}}}{m_\phi}\cos\left(\omega_\phi (t+L))- m_\phi\vec\beta\cdot\vec L\right)  - \kappa \frac{\sqrt{2 \rho_{\rm DM}}}{m_\phi} \cos\left(\omega_\phi t-m_\phi \vec\beta\cdot\vec 0\right) \\ \nonumber
&=& -2 \kappa \frac{\sqrt{2 \rho_{\rm DM}}}{m_{\phi}} \sin\left( \frac{m_{\phi} L (1- \beta) }{2}\right) \sin \left(m_{\phi}\left[t + \frac{L (1- \beta) }{2}\right]\right).
\end{eqnarray}
In a Mössbauer measurement, the tiny energy shift we predict can be resolved with exquisite precision.

%However, the expression in Eq.~\eqref{eq:Energy_Shift} holds only when the dark‑matter oscillation frequency $m_{\phi}$ does not exceed the nuclear transition’s natural linewidth $\Gamma$~\cite{Banerjee:2024bkp}.  Physically, the nucleus requires a time on the order of $1/\Gamma$ to emit its photon; if $m_{\phi} \gg  \Gamma$, the rapid field oscillations no longer produce a uniform shift but instead generate sidebands around the resonance. We are interested in the regime where $m_{\phi} \lesssim  \Gamma$, since in this case the entire Mössbauer line shifts coherently. In this limit, the observed energy shift oscillates at exactly the frequency $m_{\phi}$, providing a striking and unambiguous signature of ultra‑light dark matter.

While $^{57}\text{Fe}$ is currently the most widely used nucleus in Mössbauer spectroscopy, a number of other isotopes are known to exhibit even sharper resonance lines~\cite{MEDC}. Among them, the 93-keV transition in $^{67}\text{Zn}$, with a fractional linewidth of $\Gamma_0 / E_0 \sim 10^{-15}$~\cite{PhysRevB.15.3291}, has been successfully employed in high-resolution Mössbauer spectroscopy, particularly in gravitational redshift experiments~\cite{KATILA198151, Potzel1992}. This demonstrates the potential of narrow-linewidth nuclear transitions for precision measurements.

Motivated by this, isotopes with even narrower intrinsic linewidths have attracted increasing attention. For instance, the $E_0=88$ keV transition in $^{109}$Ag has an extremely small natural width $\Gamma_0$ and can, in principle, achieve an extraordinarily small fractional linewidth of $\Gamma_0/E_0\sim 10^{-22}$. This enables unprecedented spectroscopic resolution and has been proposed for detecting gravitational effects at small scales~\cite{Gao:2023ggo}.

Another particularly promising candidate is $^{45}\text{Sc}$, which exhibits a nuclear transition at $12.4~\text{keV}$ with an isomeric lifetime of $\tau_0 = 0.47~\text{s}$ and a natural linewidth of $\Gamma_0 = \hbar/\tau_0 = 1.4~\text{feV}$, corresponding to an extraordinarily high quality factor $Q = E_0/\Gamma_0 \simeq 10^{19}$. However, unlike some other Mössbauer isotopes, the application of $^{45}\text{Sc}$ is limited by the relatively small branching ratio of the $\beta^-$ decay of its parent nucleus $^{45}\mathrm{Ca}$ into the excited state relevant for the Mössbauer transition, which suppresses the achievable resonant photon flux from radioactive sources. 

Earlier attempts to achieve resonant excitation of $^{45}\text{Sc}$ at third-generation synchrotron radiation facilities were unsuccessful due to insufficient spectral flux. This limitation has only recently been overcome with the advent of narrow-band, high-repetition-rate X-ray free-electron laser (XFEL), most notably the European XFEL~\cite{Shvydko2023,Decking:2020bct,liu2023cascaded}, where resonant excitation of $^{45}\text{Sc}$ has been demonstrated. Next-generation facilities, such as the Shanghai High Repetition Rate XFEL and Extreme Light Facility (SHINE)~\cite{Wan:2022het,Zhao:2017ood,ZHAO2016720} and the Shenzhen Superconducting Soft X-ray Free Electron Laser ($\mathrm{S^{3}FEL}$)~\cite{Wang:2023vzd,Li:2024zus}, are currently under construction and are expected to provide high-brightness, high-repetition-rate X-ray beams suitable for such resonant nuclear excitation.

In this work, we focus on $^{109}\text{Ag}$ and $^{45}\text{Sc}$ as benchmark isotopes to investigate the projected sensitivity of Mössbauer measurements to shifts of nuclear transition energies induced by ultralight dark matter through recoil-free photons. Key nuclear parameters of these isotopes are summarized in Table~\ref{tab:isotope}, including the transition energy ($E_0$), natural linewidth ($\Gamma_0$), fractional linewidth ($\Gamma_0 / E_0$), natural lifetime, experimentally achieved resonance width ($\Gamma_\text{exp}$), and the radioactive parent nuclear lifetime.
\begin{table}[ht]
\centering
\begin{tabular}{c|c|c|c|c|c|c}
\hline\hline
Isotope & \makecell{Nuclear transition \\ energy $\mathrm{E_{0}}$ (keV)} & Natural Width $\Gamma_{0}$ (eV) & $\Gamma_{0}/\mathrm{E_{0}}$ & Lifetime &$\Gamma_{\mathrm{exp}}$ (eV) & Parent lifetime (day) \\
\hline
$^{109}\text{Ag}$ & 88 & $2.3\times 10^{-17}$ & $1.2\times 10^{-22}$ & 27.6~$\mathrm{s}$ & $1.9\times 10^{-16}$ & 461 ($^{109}$Cd) \\
\hline
$^{45}\text{Sc}$  & 12.4 & $1.4\times 10^{-15}$ & $1.1\times 10^{-19}$ & 0.47~$\mathrm{s}$ & $7.0\times 10^{-13}$~\cite{Liu:2025ilx} & 163 ($^{45}$Ca)$^*$ \\
%\hline
%$^{67}\text{Zn}$  & 93.3 & $5.0\times 10^{-11}$ & $5.4\times 10^{-16}$ & 6.4~$\mathrm{\mu s}$ & $7.5\times 10^{-11}$~\cite{Gratta:2020hyp} & 3.3 ($^{67}$Ga)\\
\hline\hline
\end{tabular}
\caption{Properties of selected M\"ossbauer isotopes $^{109}\text{Ag}$ and $^{45}\text{Sc}$, including nuclear transition energy $\mathrm{E_0}$, natural linewidth $\Gamma_{0}$, the ratio $\Gamma_{0}/\mathrm{E_0}$, natural lifetime, experimentally achived resonance width $\Gamma_\text{exp}$, and the radioactive parent nucleus lifetime.  
The experimentally achieved resonance widths for $^{45}$Sc is adopted from observations. 
For $^{109}$Ag, we assume 1.9$\times 10^{-16}$ eV for an estimate~\cite{Gao:2023ggo}. 
$^{*}$ Spontaneous $^{45}\mathrm{Ca} \to ^{45}\mathrm{Sc}$ decay has a major branching fraction 
suppression ${\cal O}(10^{-5})$ to the first $^{45}$Sc excited state, 
and here we list $^{45}$Sc 
as a candidate due to recent success 
at XFEL.}
\label{tab:isotope}
\end{table}

\section{The STATIONARY SCHEME}
\label{sec:Stationary_Scheme}

\begin{figure}[htbp]
\centering
\includegraphics[width=0.6\textwidth]{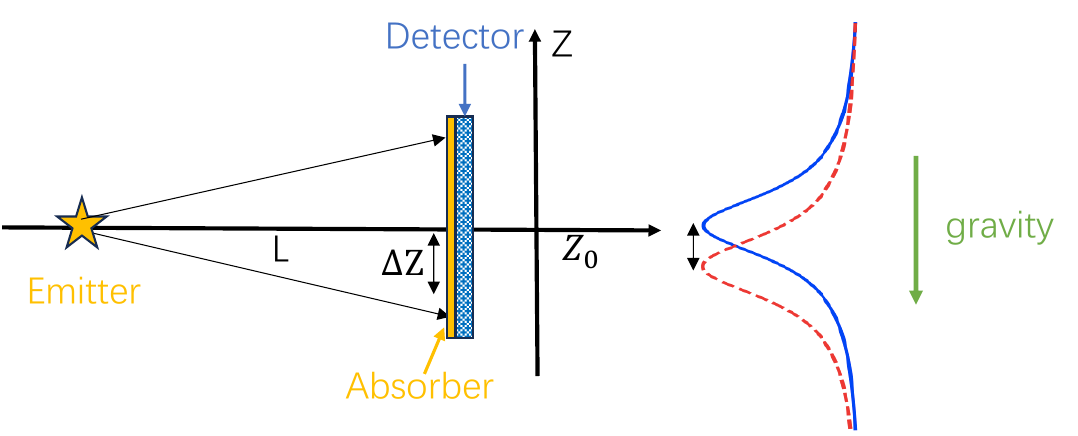}
\caption{Schematic of the static Mössbauer measurement for detecting dark-matter-induced energy shifts. The emitter and absorber are separated by a fixed baseline $L$, while the absorber is vertically displaced by $\Delta Z$ to compensate for the energy mismatch via gravitational redshift. The absorption spectrum measured by the detector is also shown schematically, where the shift of the resonance peak is used to infer the corresponding energy offset $\Delta E$.}
\label{Fig:Measurement}
\end{figure}

To probe possible variations in nuclear transition energies induced by an oscillating dark matter background, we consider a static measurement scheme based on a Mössbauer resonance absorber with an ultra-narrow linewidth~\cite{Gao:2023ggo}. As illustrated in Fig.~\ref{Fig:Measurement}, a fixed baseline $L$ is established between the emitter and the absorber, while the absorber and the detector are arranged along the vertical direction. The right panel schematically displays the absorption spectrum measured by the detector. In the presence of the Earth's gravitational field, the displacement of the absorption peak can be used to infer the nuclear energy shift.

Under standard Mössbauer resonance conditions, when the emitter and the absorber are located at the same gravitational potential, i.e., at the same height $Z_0$, the nuclear transition energies of the source and the absorber are perfectly matched, allowing the emitted photons to be resonantly absorbed without recoil. If dark matter couples to nucleons, it can induce tiny shifts in the nuclear energy levels, leading to a small energy offset $\Delta E$ in the emitted photons. In this case, an absorber placed at the same height no longer satisfies the resonance condition, and resonant absorption is suppressed, so that more photons pass through the absorber and are recorded by the detector.

Resonance can be restored by exploiting the gravitational redshift effect to compensate for the energy mismatch. In the Earth's gravitational field, the photon energy changes as it propagates along the vertical direction. Therefore, by introducing a small vertical displacement $\Delta Z$ of the absorber, one creates a difference in gravitational potential that induces a corresponding redshift or blueshift in the photon energy. In the weak-field limit near the Earth's surface, the relation between the energy shift and the vertical displacement is given by
\begin{equation}
\frac{\Delta E}{E} \simeq \frac{g \Delta Z}{c^{2}},
\label{eq: delta_E_Z}
\end{equation}	
where $g$ is the gravitational acceleration and $E$ is the photon energy. When the gravitationally induced shift compensates the dark-matter-induced energy offset $\Delta E$, resonance is restored at the displaced position. Thus, by measuring the vertical displacement $\Delta Z$ at which resonant absorption is maximized, for example through scanning the absorption line, one can infer the corresponding energy shift $\Delta E$, thereby providing a means to probe or constrain the coupling between dark matter and nuclear energy levels.

Let us assume that the emitter is stationary at a vertical position $Z_{S}$. The emission line of the source is narrow and unsplit, with a natural linewidth $\Gamma$. The measured spectral lineshape is Lorentzian, centered at a resonance energy $E_{0}$. Denoting the total photon emission rate of the source as $N_{0}$, the differential number of recoil-free photons with energy $E_{\gamma}$ emitted per unit energy and time is given by~\cite{gutlich2011mossbauer,Gao:2023ggo}:
\begin{eqnarray}
N_{\rm RF}(E) = N_{0} f_{S}\cdot  \frac {\Gamma /2\pi} {[E-E_0]^2 + (\Gamma /2)^2 }~,~~
%\frac{{\rm d}N_{\rm RF}(E)}{{\rm d}E ~{\rm d} t} = N_{0} f_{S}\cdot  \frac {\Gamma /2\pi} {[E-E_0]^2 + (\Gamma /2)^2 }~,~~
\label{eq:p_emmit}
\end{eqnarray}
where $f_{S}$ is the recoil-free fraction of the source emission.

To achieve the desired performance in a static configuration, the detector must have a good spatial resolution, while the finite sizes of the absorber and detector should allow them to cover a narrow vertical range. Within this configuration, the Mössbauer transmission function can be formulated by replacing the conventional Doppler shift with a gravitationally induced energy correction arising from the vertical displacement.
The transmission function C(Z) at a detector height Z is:
\begin{eqnarray}
\label{formula:spectrum}
C(Z)  &=&  N_0 ~{e}^{-\mu_e t'} \cdot \left[(1-f_S) + \vphantom{\frac{XX}{XX}}\right. \left.\int_{-\infty}^{\infty} f_S\xi(Z_S,E_0)\cdot {\rm e}^{-t \xi(Z,E_0+\Delta E_0)\Gamma /2\pi } {\rm d}E \right], \label{eq:a_emmit} \\
{\rm}~\xi(Z,E_0)&\equiv& \frac{\Gamma /2\pi}{[E-g(Z-Z_S)E-E_0]^2 + (\Gamma /2)^2}. \nonumber
\end{eqnarray}
Here, $\Delta E_{0}$ represents the intrinsic energy offset between the emitter and absorber, which can be compensated by a small vertical height difference within the range of the detector. At resonance, this correction satisfies 
$g \left< Z - Z_{S}\right>E = - \Delta E_{0}$, and we define the corresponding central detection height as $Z_{0} = \left<Z\right>$ for convenience.
%We adopt natural units with $\hbar=c=1$. 
The attenuation factor $e^{-\mu_{e} t^{\prime}}$ accounts for non-resonant absorption, where $t^{\prime}$ is the absorber thickness expressed as the area density, i.e., $t'=\rho L$, with units of $\mathrm{g\, cm^{-2}}$~\cite{gutlich2011mossbauer},
and $\mu_{e}$ is the mass absorption coefficient at resonance. The effective resonant optical depth is 
$t = f_{A} N_{M} \sigma_{0}$, where $f_{A}$ is the recoil-free fraction at the absorber, $N_{M}$ is the number of Mössbauer nuclei per unit area (which grows with absorber thickness), and 
$\sigma_{0}$ is the peak resonant cross section.

For practical calculations, the transmission function can be approximated by a simplified parameterization that highlights the gravitational energy shift~\cite{Gao:2023ggo}:
\begin{eqnarray}
C(Z)= N_{0} ~{e}^{-\mu_e t'} \left\{1-f_S~\epsilon\cdot\frac{ \Gamma_{\rm exp}^2}{[g(Z-Z_0)E_0]^2+\Gamma_{\rm exp}^2}\right\}.
\label{eq:parametrization}
\end{eqnarray}
Here, $\epsilon$ is the effective resonant absorption fraction, and $\Gamma_{\text{exp}}$ is the observed linewidth after transmission through the absorber. Both quantities depend nontrivially on the thickness of the absorber. The quantity $f_S$ represents the fraction of recoil-free emission, while $f_S \epsilon$ represents the total recoil-free absorption fraction. 
% The term $1 - f_S$ denotes the fraction of emission that involves recoil.

We consider a factor that induces a frequency shift $\Delta f(t)$ in the resonant absorption of photons by the absorber. This shift results in a vertical displacement of the resonance position, given by
\begin{eqnarray}
Z_0 \rightarrow Z_0(t) = Z_0 + g^{-1} \frac{\Delta f(t)}{f_\gamma}.
\label{eq:h_movement}
\end{eqnarray}
We define the displacement as
$\Delta Z_{0}(t) \equiv Z_{0}(t) - Z_{0}$.
This spatial shift can be inferred from variations in Mössbauer absorption efficiency at the detector. If the spatial resolution is sufficiently high,
both $\Delta Z_{0}(t)$ and the corresponding $\Delta f(t)$ can be measured with a sensitivity exceeding the effective Mössbauer linewidth, $\Gamma_{\mathrm{exp}}/E_{0}$.

\section{Dark Matter Signal}
\label{sec:DM_signal}

The interaction between oscillating dark matter and atomic nuclei can induce tiny shifts in nuclear energy levels. This, in turn, causes a slight change in the resonance absorption energy observed in the Mössbauer effect, leading to a vertical displacement of the absorber's resonance absorption position, denoted as $Z_{0}(t)- Z_{0}$. This vertical displacement, $\Delta Z_{0}$, is quantitatively related to the shift in resonance energy $\Delta E$ through the following relation:
\begin{eqnarray}
\Delta Z_{0} \equiv Z_{0}(t) - Z_{0} =  g^{-1} \frac{\Delta f(t)}{f_\gamma} =  g^{-1} \frac{\Delta E}{E_{0}}.
\label{eq:delta_h_delta_E}
\end{eqnarray}

The $88~\mathrm{keV}$ emission line of $^{109}\text{Ag}$ has a natural linewidth of $2.3 \times 10^{-17}~\mathrm{eV}$~\cite{Gratta:2020hyp,MEDC}.
When both emission and absorption processes are considered, the ideal resonance width effectively doubles. 
%In practice, the observed linewidth is widened due to smearing effects in the material of the sample, which include the impact of the effective absorber thickness~\cite{MARGULIES1961131,Cranshaw_1974,shenoy1974}. For an effective absorption depth of $t = 8$, this broadening results in a factor of 4.1 increase. Under such conditions, the measured linewidth for the 88 keV transition becomes approximately $\Gamma_{\text{exp}} = 1.9 \times 10^{-16}~\mathrm{eV}$. 
This $^{109}\text{Ag}$ linewidth corresponds to a vertical displacement of about 
$20~\mu \mathrm{m}$ in Earth's gravitational field with 
$g = 9.8~\mathrm{m/s^2}$. To ensure precise localization, the detector is designed with a spatial resolution of 
$10~\mu \mathrm{m}$, which is half the broadened absorption peak width. We further assume that the detector maintains a high efficiency, approaching 100\%, for detecting these high-energy X-ray photons.

The strength of the resonance is inferred by measuring the transmitted (unabsorbed) photon flux through an absorber layer as a function of the vertical position (Z). The absorption fraction is given by Eq.~\ref{eq:parametrization}, which predicts a position-dependent suppression of the photon flux near the resonance.
To quantify this effect, we discretize the detector along the vertical direction into bins of width $\delta Z$. The detector thus measures a binned spatial distribution of photon counts. We denote by $C_i(Z_i; Z_0)$ the expected number of photon counts in the (i)-th bin centered at position $Z_i$, where $Z_0$ specifies the location of the resonance peak. Here $Z_i$ denotes the fixed central position of the (i)-th bin and is not a free parameter in the fit. Far from the resonance region, the absorption becomes negligible, and the count rate approaches an asymptotic value $C_\infty$, corresponding to the unperturbed photon flux.

For a given Mössbauer source, the bin width is chosen to match the characteristic spatial scale over which the resonance varies due to gravitational energy shifts. Specifically, we set
$\delta Z \sim \frac{\Gamma_{\mathrm{exp}}}{g E_0}$,
and adopt $\delta Z/2 = 10~\mu\mathrm{m}$ for $^{109}\text{Ag}$, which ensures that the resonance profile is adequately resolved without introducing unnecessary statistical fluctuations.

In the Monte Carlo simulation, the simulated data $C_i^{\rm exp}$ are generated by drawing from a Poisson distribution with mean given by the theoretical expectation $C_i(Z_i; Z_0^{\rm true})$. We set $Z_0^{\rm true}=0$, which defines the origin of the spatial coordinate. Since the model depends only on the relative displacement $Z - Z_0$, this choice does not affect the statistical uncertainty but allows us to directly interpret the fitted value as the deviation from the true resonance position.

Since the expected counts $C_i(Z_i; Z_0)$ depend on the resonance position $Z_0$, we determine the best-fit value of $Z_0$ by minimizing the following $\chi^2$ function:
\begin{eqnarray}
\chi^2(Z_0) = \sum_{i}\frac{\big[C_i(Z_i; Z_0) - C_i^{\rm exp}\big]^2}{\Delta_i^2},
\label{eq:chi_square}
\end{eqnarray}
where $C_i^{\rm exp}$ denotes the observed number of counts in the (i)-th bin. In this analysis, $Z_0$ is treated as the only free parameter, while all other quantities are fixed. We emphasize that the above $\chi^2$ definition does not include detector effects or systematic uncertainties (e.g., via nuisance or pull terms), and thus corresponds to an idealized, statistics-dominated scenario.

The above $\chi^2$ definition corresponds to a maximum-likelihood estimator under the assumption that the statistical fluctuations in each bin are approximately Gaussian. This approximation is justified because the photon counts in each bin follow a Poisson distribution, and for sufficiently large counts, the Poisson distribution approaches a Gaussian with variance equal to its mean. Accordingly, in the statistics-dominated regime we take
$\Delta_i = \sqrt{C_i}$,
where $C_i$ represents the expected counts in the (i)-th bin.

The spatial resolution on the resonance position $Z_0$ is then determined from the curvature of the likelihood function (or equivalently the $\chi^2$ function) around its minimum. In practice, this corresponds to the statistical uncertainty on the fitted parameter $Z_0$, which can be extracted either from the second derivative of $\chi^2$ or from repeated Monte Carlo realizations of the experiment.

\begin{figure}[htbp]
\centering
\includegraphics[width=0.6\textwidth]{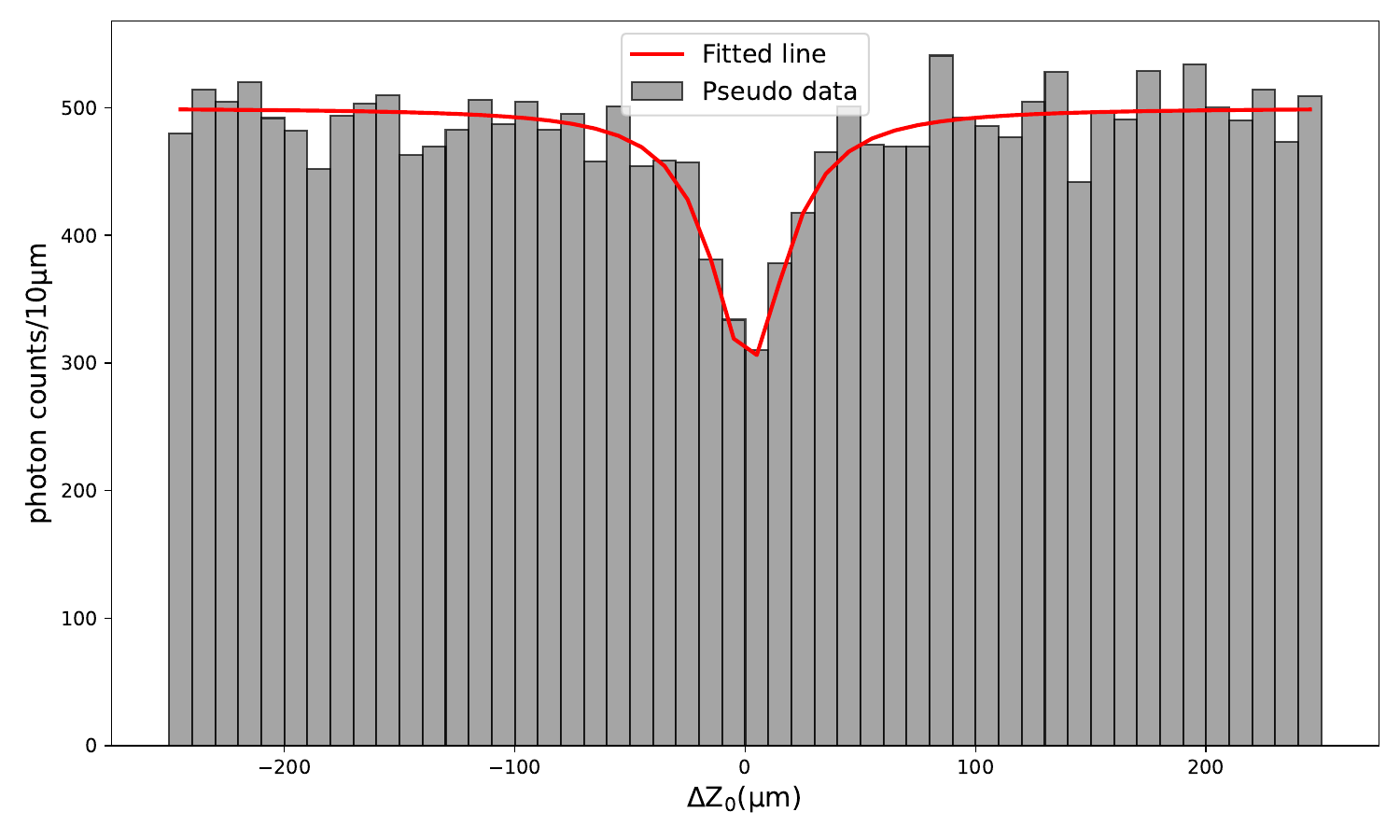}
\caption{For the $^{109}\text{Ag}$ nucleus: Simulated peak position accuracy $\Delta Z_{0}$	obtained from a single pseudo-experiment measuring the position of the absorption Lorentzian peak, assuming an experimental linewidth of 
$20~\mathrm{\mu m}$, z-axis bin width of $10~\mathrm{\mu m}$, 500 arriving photons per bin, recoil-free fraction $f_S$ =0.5, and absorption fraction $\epsilon$=0.8. The value $f_S$ =0.5 is an unphysical value adopted solely to enhance the visibility of the absorption profile in this illustrative simulation.}
\label{Fig:Ag109_simulated}
\end{figure}

The simulated position resolution of the Lorentzian peak is shown in Fig.~\ref{Fig:Ag109_simulated}.
For $^{109}\mathrm{Ag}$, the experimental linewidth is $20~\mu m$, and the bin width along the z-axis is $10~\mu m$. Each bin assumes a number of incident 88 keV photons determined by the intensity and distance of the source. The recoil-free fraction is set to $f_{S} = 0.5$, and the absorption fraction is taken to be $\epsilon = 0.8$, corresponding to a near-optimal absorber configuration~\cite{Gao:2023ggo}.
The photon flux measured in the resonance bins decreases due to resonance absorption, as shown by the simulated photon counts in Fig.~\ref{Fig:Ag109_simulated}. We note that the value $f_{S}=0.5$ is an unphysical value adopted solely for illustrative purposes, to enhance the visibility of the absorption profile in this simulation. For the final sensitivity estimates presented in this work, we use the experimentally measured value
$f_{S}=0.05$ for $^{109}\mathrm{Ag}$.

For our analysis, we assume that the absorbers and detectors are uniformly arranged on a circle, with the radioactive emitter placed at the geometric center. 
The angular coverage $\theta$ corresponding to the vertical height difference $\Delta Z_{0}$ of the resonant absorption peaks on the detector is given by 
\begin{eqnarray}
\theta = \arctan \left(\frac{\Delta Z_{0}}{L} \right),
\label{eq:theta}
\end{eqnarray}
where $L$ is the distance between the source and the absorber. In our setup, we take $L = 1~m$. For a circular detector geometry, the fraction of the total solid angle subtended by the detector is given by
\begin{eqnarray}
f_{\theta} = \frac{2\pi \theta}{4\pi} = \frac{\theta}{2}.
\end{eqnarray}
This quantity represents the effective detection solid-angle coverage normalized to the full $4\pi$ steradians.

We consider a radioactive source with an activity of $A= 0.1~\mathrm{Ci}$ ( corresponding to $3.7~\mathrm{GBq}$). The total number of emitted photons during the measurement is
\begin{eqnarray}
N_{total} = A \times P_{\gamma} \times f_{\mathrm{BR}} \times \Delta t \times N_{\rm{exp}}, 
\label{eq:N_total_gamma}
\end{eqnarray}
where $P_{\gamma} = 1/(1 + \alpha_{\mathrm{T}})$ is the probability of photon emission, $\alpha_{\mathrm{T}}$ is the internal conversion coefficient, defined as the ratio between the number of internal conversion electrons $N_{\mathrm{IC}}$ and the number of gamma emissions $N_{\gamma}$,
$f_{\mathrm{BR}}$ denotes the branching ratio for the transition of interest,
$\Delta t = T/3$ is the integration time per data segment, with $T = 2\pi/m_{\phi}$ being the oscillation period associated with dark matter of mass $m_{\phi}$. The choice $\Delta t = T/3$ represents a compromise between time resolution and photon statistics. A significantly longer integration time would average out the oscillatory signal, while a much shorter one would reduce the photon counts per bin and thus degrade the statistical sensitivity. This choice ensures that the modulation induced by dark matter oscillations remains resolvable without being washed out. The factor $N_{\rm exp} = 10^{6}$ accounts for the statistical enhancement arising from the number of dark matter oscillation periods within its coherence time scale. Here, the coherence time is taken to correspond to a single data-taking period.
We will take the DM coherent time scale $t_c=N_{\rm{exp}}\cdot 2\pi m_\phi^{-1}$ as the benchmark coherent measurement time in this sensitivity estimate, and longer time data-taking would be considered as repeated yet independent measurements.   
For each energy bin, the number of photons contributing to that bin is 
\begin{eqnarray}
N_\gamma = N_{total} \times f_{\theta}.
\end{eqnarray}
This framework provides a quantitative estimate of the photon statistics available per bin under realistic experimental conditions, which is crucial for determining the achievable sensitivity of the measurement.

For $^{45}\text{Sc}$, two approaches can be employed to generate M\"ossbauer radiation. The conventional route assumes a radioactive $^{45}\text{Ca}$ source with an activity of 0.1~Ci. However, directly using $^{45}$Ca is inefficient as they almost always decay to the ground state of $^{45}$Sc. An alternative approach relies on synchrotron radiation to resonantly excite the $^{45}\text{Sc}$ nucleus, for instance, by a high-intensity X-ray beam~\cite{shvydko2023resonant}. The resonance excitation of $^{45}$Sc by XFEL beam generates coherently scattered photons in the forward direction~\cite{sturhahn2004nuclear}.
The yield of pure nuclear-resonant photons can be estimated from the incident photon flux within the narrow resonance bandwidth. Owing to the relatively long lifetime of $^{45}$Sc (0.47~s), the nuclear-resonant signal can be effectively distinguished from the directly transmitted incident beam.
%The pure nuclear resonance photons can be estimated with the incident photon intensity within the narrow bandwidth. The lifetime of $^{45}$Sc is 0.47s, thus allowing for distinguishing the nuclear resonant beam from the direct incident beam.
The beam-excited M\"ossbauer resonance photon flux $N_{\gamma}^{\mathrm{MS}}$ can be estimated as
\begin{eqnarray}
N_{\gamma} ^{\mathrm{MS}} = N_{\gamma}^{\mathrm{SR}} \frac{\Gamma } {\Delta E^{\text{SR}}},
\label{eq:N_gamma_MS_SR}
\end{eqnarray}
where $N_{\gamma }^{\mathrm{SR}}$ denotes the spectral photon flux of the synchrotron radiation, $\Gamma$ is the natural linewidth of the 
$^{45}\text{Sc}$ nuclear transition, and $\Delta E^{\mathrm{SR}}$ represents the energy resolution of the synchrotron source.

The resonant photons are emitted within a narrow angular distribution and, in principle, can be efficiently collected by the absorber or detector. In addition, due to the coherent forward scattering induced by synchrotron excitation, the emitted photons are strongly collimated in the forward direction. In our setup, detectors are placed along the forward direction, and optical elements can be employed to adjust the beam spot size to match the detector acceptance. Therefore, for the purpose of sensitivity estimation, we assume an idealized configuration in which the majority of Mössbauer photons produced via synchrotron radiation are received by the absorber or detector.

To estimate the photon statistics at the absorber, we approximate the transverse beam profile as uniform over an effective illuminated area. The number of photons intercepted by the absorber is then given by the ratio of the detector area to the beam spot area.
Specifically, the absorber acceptance can be written as
\begin{equation}
f_{\mathrm{geo}} = \frac{A_{\mathrm{abs}}}{A_\mathrm{beam}} = \frac{h x}{\pi r^{2}} \, ,   
\end{equation}	
where $h$ and $x$ denote the vertical and horizontal dimensions of the absorber or detector, respectively, and $R$ is the effective beam spot radius at the detector location.
In our setup, we consider a vertical separation scale 
$h \approx \Delta Z= 0.26~\mathrm{m}$ and a detector width 
$x= 0.1 ~\mathrm{m}$. To ensure full coverage of the vertically distributed absorbers/detectors, we conservatively assume a beam spot radius $R \approx \Delta Z$. Under this assumption, the geometrical acceptance reduces to
\begin{equation}
f_{\mathrm{geo}} \approx \frac{x} {\pi \Delta Z} \, .
\end{equation}
Accordingly, the number of photons received by the absorber can be estimated as
\begin{eqnarray}
N_{\gamma } = N_{\gamma }^{\text{MS}}\times f_{\mathrm{geo}} \times  P_{\gamma} \times \Delta t \times N_{\text{exp}}.
\end{eqnarray}
We note that this estimate corresponds to an idealized configuration with a uniform beam profile. In practice, the beam profile can be tailored using X-ray optics to control its spatial distribution and divergence. For instance, defocused illumination or beam-shaping techniques can be employed to modify the beam size and improve the overlap with the detector geometry, thereby enhancing the effective photon collection efficiency ~\cite{Zhang:ok5151,10.1117/12.2061941}.
In our analysis, we adopt an effective beam energy resolution of $\Delta E^{\mathrm{SR}} = 1~\mathrm{eV}$ for a modern synchrotron radiation source, such as the European XFEL~\cite{Shvydko2023, Decking:2020bct, liu2023cascaded}, and a spectral photon flux of $N_{\gamma}^{\mathrm{SR}} = 5 \times 10^{14} \, \mathrm{ph\,s^{-1}\,eV^{-1}}$~\cite{Shvydko2023}.

To ensure that the signal-to-noise ratio reaches the level of at least $3\sigma$, the numbers of detected signal and background events must satisfy:
\begin{eqnarray}
\frac{N_{S}}{\sqrt{N_{S} + N_{B}}} &\geq& 3\sigma,
\label{eq:N_SB_Ratio}
\end{eqnarray}
where $N_{S}$ denotes the number of signal events and is given by $N_{S} = \epsilon f_{S} N_{\gamma }$. The remaining events are treated as background, so the background count is $N_{B} = N_{\gamma} - N_{S}$. Substituting these expressions into Eq. \eqref{eq:N_SB_Ratio} and solving for the total number of detected photons $N_{\gamma}$, we obtain the minimal number of photons required to achieve a $3\sigma$ detection significance:
\begin{eqnarray}
N_{\gamma } &\geq& \left ( \frac{3\sigma }{\epsilon f_{S} } \right )^{2}.
\label{eq:N_minimal_gamma}
\end{eqnarray}
This condition sets a lower bound on the total number of photons that must be collected in order for the signal to be statistically distinguishable from the background at the desired confidence level.

%Generally, one needs to balance a larger absorption fraction against the mass attenuation. Reference~\cite{Long1983} showed that a mass attenuation near $\mu_{e} t^{\prime} \approx 2$ provides the best sensitivity, with a corresponding absorber efficiency of $\epsilon \approx 0.8$~\cite{Gao:2023ggo}. For $^{109}\text{Ag}$ recoil-free fraction $f_{S} =0.05$~\cite{Wildner1979ANA}, and the absorption fraction $\epsilon=0.8$. We use the $\chi^{2}$ formula Eq.~\ref{eq:chi_square}  to determine the fitted $\Delta Z_0$ value. The fit for one simulation is presented in Fig.~\ref{Fig:Ag109_simulated}. We then perform 1000 simulation runs to obtain the 95\% confidence interval for $\Delta Z_0$. We use the $\chi^{2}$ formula Eq.~\ref{eq:chi_square} to determine the fitted $\Delta Z_0$ value. For each simulated dataset, the $\chi^{2}$ is constructed by comparing the expected counting rate as a function of (Z) with the simulated data, taking into account the statistical uncertainty of each data point. The best-fit value of $\Delta Z_0$ is then obtained through a one-parameter minimization of the $\chi^{2}$function. The fit for one representative simulation is shown in Fig.~\ref{Fig:Ag109_simulated}. We then perform 1000 independent simulation runs to obtain the 95\% confidence interval for $\Delta Z_0$, which is derived from the distribution of the fitted values across all realizations.

Generally, one needs to balance a larger absorption fraction against mass attenuation. Ref.~\cite{Long1983} showed that a mass attenuation near $\mu_{e} t^{\prime} \approx 2$ provides the optimal sensitivity, corresponding to an absorber efficiency of $\epsilon \approx 0.8$~\cite{Gao:2023ggo}. For $^{109}\text{Ag}$, we adopt a recoil-free fraction $f_{S} = 0.05$~\cite{Wildner1979ANA} and an absorption fraction $\epsilon = 0.8$.
We determine the fitted $\Delta Z_0$ using the $\chi^{2}$ function defined in Eq.~\ref{eq:chi_square}. For each simulated dataset, $\chi^{2}$ is constructed by comparing the expected counting rate as a function of $Z$ with the mock data, taking into account the statistical uncertainty in each bin. The best-fit value of $\Delta Z_0$ is then obtained by minimizing the $\chi^{2}$ function with respect to this single parameter. An example fit from one representative simulation is shown in Fig.~\ref{Fig:Ag109_simulated}.
To estimate the statistical uncertainty, we perform 1000 independent Monte Carlo realizations and extract the 95\% confidence interval for $\Delta Z_0$ from the distribution of the best-fit values across all simulations.

The resulting 95\% confidence interval on $\Delta Z_0$ is then translated into the corresponding shift in the resonance absorption energy using Eq.~\ref{eq:delta_h_delta_E}, which encodes the dark matter-induced modulation of the resonance condition. Combining this with the theoretical relations in Eqs.~(\ref{eq:delta_E_gamma}), (\ref{eq:delta_E_quark}), and (\ref{eq:delta_E_gluon}), we derive constraints on the couplings between dark matter and Standard Model particles, including photons, quarks, and gluons. This procedure establishes a direct link between the experimentally accessible observable and the underlying dark matter parameters, allowing the projected sensitivity to be interpreted as bounds on dark matter interactions.

%We extend the methodology previously applied to the $^{109}\text{Ag}$  nucleus to additional emitter: $^{45}\text{Sc}$. For $^{45}\text{Sc}$, the $12.4~\mathrm{keV}$ nuclear transition has an intrinsic experiment linewidth of $7.0 \times 10^{-13}~\mathrm{eV}$~\cite{Liu:2025ilx}, which corresponds to a vertical energy shift of approximately $\delta Z = 0.52~\mathrm{m}$ in Earth's gravitational field. To resolve this gravitationally induced broadening, the detector is designed with a spatial resolution of $0.26~\mathrm{m}$, i.e., half the broadened absorption width. For $^{67}\text{Zn}$, the $93.3~\mathrm{keV}$ transition has a much larger experiment linewidth of $7.5 \times 10^{-11}~\mathrm{eV}$~\cite{Gratta:2020hyp}, resulting in a gravitational broadening of $\delta Z = 7.4~\mathrm{m}$; accordingly, the detector spatial resolution is set to $3.7~\mathrm{m}$. The Lamb–Mössbauer factors at room temperature, estimated within the Debye approximation, are approximately 0.75~\cite{Shvydko2023} for $^{45}\text{Sc}$ and 0.2~\cite{Potzel1993} for $^{67}\text{Zn}$, respectively. In both cases, we employ a $\chi^{2}$-minimization procedure to extract the best-fit vertical displacement $Z_{0}$. Based on 1000 Monte Carlo simulations, we determine the 95\% confidence intervals of $\Delta Z_0$ for $^{45}\text{Sc}$ and $^{67}\text{Zn}$.

We extend the methodology previously applied to the $^{109}\text{Ag}$ nucleus to an additional emitter $^{45}\text{Sc}$. For $^{45}\text{Sc}$, the $12.4~\mathrm{keV}$ nuclear transition has an experimentally achieved linewidth of $7.0 \times 10^{-13}~\mathrm{eV}$~\cite{Liu:2025ilx}, which corresponds to a vertical energy shift of approximately $\delta Z = 0.52~\mathrm{m}$ in Earth's gravitational field. To resolve this gravitationally induced broadening, the detector is designed with a spatial resolution of $0.26~\mathrm{m}$, i.e., half of the broadened absorption width.
The Lamb–Mössbauer factor at room temperature, estimated within the Debye approximation, is approximately 0.75~\cite{Shvydko2023} for $^{45}\text{Sc}$. In this case, we employ a $\chi^{2}$-minimization procedure to extract the best-fit vertical displacement $Z_{0}$. Based on 1000 Monte Carlo simulations, we determine the 95\% confidence interval of $\Delta Z_0$.

In Table~\ref{tab:isotope_parameters}, we summarizes the key nuclear parameters of these isotopes, including internal conversion coefficient  $\alpha_\mathrm{T}$~\cite{Internal_Conversion_Coeff}, probability of gamma $P_{\gamma}$, branching ratio $f_{\mathrm{BR}}$~\cite{Internal_Conversion_Coeff}, recoil-free fraction $f_{S}$, and the absorption fraction $\epsilon$ is taken to be 0.8.
\begin{table}[ht]
\centering
\begin{tabular}{c|c|c|c|c}
\hline\hline
Isotope &  $\alpha_{\mathrm{T}}$ &   $P_{\gamma}$ & $f_{\mathrm{BR}}$ & $f_{S}$  \\
\hline
$^{109}\text{Ag}$ & 26.3 & 0.037 & 1 & 0.05   \\
\hline
$^{45}\text{Sc}$  & 423 & 0.0024 & $2\times 10^{-5}$ & 0.75   \\
%\hline
%$^{67}\text{Zn}$  & 0.873 & 0.534  & 0.52 & 0.2 \\
\hline\hline
\end{tabular}
\caption{Properties of selected M\"ossbauer isotopes $^{109}\text{Ag}$ and $^{45}\text{Sc}$, including internal conversion coefficient  $\alpha_\mathrm{T}$, probability of gamma $P_{\gamma}$, branching ratio $f_{\mathrm{BR}}$, recoil-free fraction $f_{S}$, and the absorption fraction $\epsilon =0.8$ throughout. }
\label{tab:isotope_parameters}
\end{table}

\section{Dark Matter - Nucleus Interactions}
\label{sec:DM_Nuc_interact}

In this section, we explore how dark matter interactions with photons, quarks, and gluons can affect the energy levels of atomic nuclei. We will present the corresponding Lagrangians for each type of interaction and derive the formulas for the resulting energy shifts in the nuclei. 

To describe the interaction between the ultra-light dark matter field $\phi$ and the Standard Model, we consider a linear coupling of $\phi$ to quarks (q), gluons ($G_{a\mu\nu}$), and photons ($F_{\mu\nu}$). The corresponding interaction Lagrangian takes the form~\cite{Gratta:2020hyp,Banerjee:2024bkp}:
\begin{eqnarray}\label{eq:lagrangian}
%\mathcal{L}\supset \sum_{\psi=u,d}\!-\frac{\phi}{f_\psi} m_\psi \bar \psi \psi - \frac{\alpha\,\phi}{4\,f_\gamma} F^2-\frac{\beta(\alpha_s)}{4\alpha_s}\frac{\phi}{f_g} G^2 \,,
%\mathcal{L}\supset  -y_{q}\phi \bar{q}q -\frac{\phi}{4\,f_\gamma} F^2 - \frac{\phi}{4 f_g} G^2 \,,
\mathcal{L}\supset \sum_{\psi=u,d}\!-\frac{\phi}{f_\psi} m_\psi \bar \psi \psi - \frac{\phi}{4\,f_\gamma} F^2 - \frac{\phi}{4 f_g} G^2 \,,
\end{eqnarray}
where, $G^2=G_{\mu\nu}^{a} G^{a\mu\nu}$ with $a$ being the color index, $F^2=F_{\mu\nu}F^{\mu\nu}$, $f_{\psi}$, $f_{\gamma}$ and $f_{g}$ denotes the interaction strength with various modules.  	

These interactions can induce time-dependent variations in fundamental constants such as the fine-structure constant $\alpha$, the down quark, and the QCD structure constant $\alpha_{s}$. Such variations lead to shifts in nuclear energy levels, which can be estimated for each type of coupling.

\textbf{Photon coupling:} The coupling $\frac{\phi}{4\,f_\gamma} F_{\mu\nu}^2$  shifts the fine structure constant  $\alpha \to \alpha - \frac{\alpha^{2} \phi}{4\,f_\gamma}$, which alters the electromagnetic self-energy of the nucleus. For a nucleus with $Z$ protons and mass number $A$, the resulting energy shift is approximately~\cite{Gratta:2020hyp, Banerjee:2024bkp}
\begin{eqnarray}
\label{eq:delta_E_gamma}
\Delta E \simeq - \frac{Z \alpha^2}{A^{1/3} r_0} \frac{\Delta \phi}{f_\gamma} \,,
\end{eqnarray}
where $r_0 \approx 1.2 \,\mathrm{fm}$ is a typical nuclear scale.

%\textbf{Quark (Yukawa) coupling:} The coupling to the light quark masses modifies the pion mass $m_\pi$ via $m_\pi^2 \propto \hat{m}$, which in turn affects the nuclear potential through one-pion exchange. The resulting energy shift is estimated as~\cite{Banerjee:2024bkp}
\textbf{Quark (Yukawa) coupling:} 
A linear Yukawa coupling between the scalar field $\phi$ and light quarks is considered in the isospin-symmetric basis, where 
$\hat{m} = (m_u + m_d)/2$. This interaction induces a fractional variation
$\Delta \hat{m} / \hat{m} = \phi / f_{\hat{m}}$, which propagates to the pion mass via $m_{\pi}^{2} \propto \hat{m}$, thereby modifying pion-mediated nuclear forces. To estimate the resulting shift in nuclear energy, the one-pion-exchange (OPE) potential between nucleons is employed. For inter-nucleon distances $r \lesssim 1/m_{\pi}$, the induced energy shift can be approximated as~\cite{Banerjee:2024bkp}
\begin{equation}
\Delta E \simeq 12.3~\mathrm{MeV} \, \frac{\Delta \phi}{f_{\hat{m}}}.
\label{eq:delta_E_quark}
\end{equation}

%As an example of the Yukawa couplings in Eq.~\ref{eq:lagrangian}, we focus on the scalar field $\phi$ coupling to the down quark through the interaction term $y_{d}\,\bar{d}d$. To estimate the resulting shift in nuclear energy levels induced by $\phi$, we consider the change in the one-pion–exchange potential and equate it to the energy shift $\Delta E$, yielding~\cite{MurayamaManyBody,Gratta:2020hyp,Banerjee:2024bkp} 
%\begin{eqnarray}
%\label{eq:delta_E_quark}
%\Delta E \simeq 2.5 \, y_{d} \Delta \phi \,.
%\end{eqnarray}

\textbf{Gluon coupling:} The coupling $\frac{\phi}{4 f_g} G_{\mu\nu}^2$ shifts the QCD structure constant $\alpha_{s} \to \alpha_{g} - \frac{ \alpha_{s} \phi}{4 f_g}$.
The induced energy shift is approximately~\cite{Gratta:2020hyp}
\begin{eqnarray}
\label{eq:delta_E_gluon}
\Delta E \simeq 0.08 \,\mathrm{GeV} \frac{\Delta \phi}{f_g} \,.
\end{eqnarray}

\section{Projected Sensitivity}
\label{sec:Projected_Senitivity}
\begin{figure}[htbp]
\centering
\includegraphics[width=0.8\textwidth]{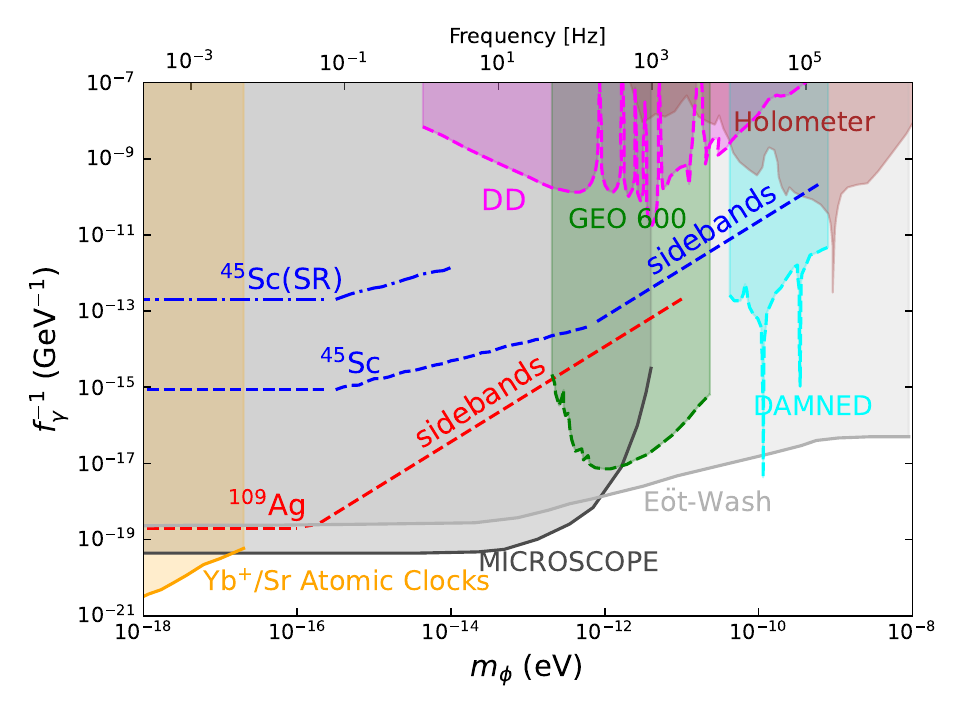}
\caption{Projected sensitivity to the scalar DM–photon coupling $f_{\gamma}^{-1}$ versus the dark matter mass $m_{\phi}$. Red and blue dashed lines denote results for $^{109}\text{Ag}$ and $^{45}\text{Sc}$, respectively. While the blue dot-dashed line corresponds to $^{45}\text{Sc(SR)}$ with synchrotron radiation.  Existing constraints from GEO 600~\cite{Vermeulen:2021epa}, DD~\cite{Aharony:2019iad}, Holometer~\cite{Aiello:2021wlp},  DAMNED~\cite{Savalle:2020vgz}, $\mathrm{Yb^{+}}$ and Sr atomic clocks~\cite{Sherrill:2023zah,Filzinger:2023zrs,Arakawa:2026mls}, as well as equivalence-principle tests by MICROSCOPE~\cite{Touboul:2017grn,Berge:2017ovy} and Eöt-Wash~\cite{Smith:1999cr,Schlamminger:2007ht}, are shown for comparison. Approximate sideband sensitivities are shown for comparison.
For $^{109}\text{Ag}$ and $^{45}\text{Sc}$ using traditional radioactive sources, the source activity is fixed at $0.1~\mathrm{Ci}$. 
For $^{45}\text{Sc}$ with synchrotron radiation (SR), a spectral photon flux of $5\times10^{14}\,\mathrm{ph\,s^{-1}\,eV^{-1}}$ is assumed. 
}
\label{Fig:f_gamma}
\end{figure}

\begin{figure}[htbp]
\centering
\begin{subfigure}{0.49\textwidth}
    \centering
    \includegraphics[width=\textwidth]{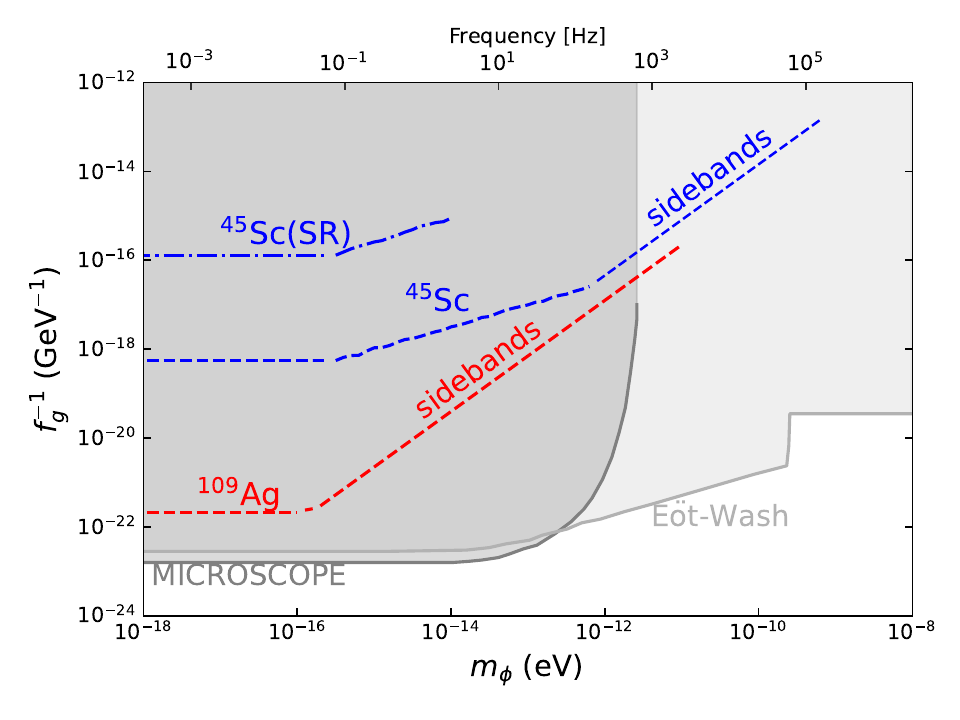}
    \subcaption{(a)}
    \label{fig:f_gluon}
\end{subfigure}
\hfill
\begin{subfigure}{0.49\textwidth}
    \centering
    \includegraphics[width=\textwidth]{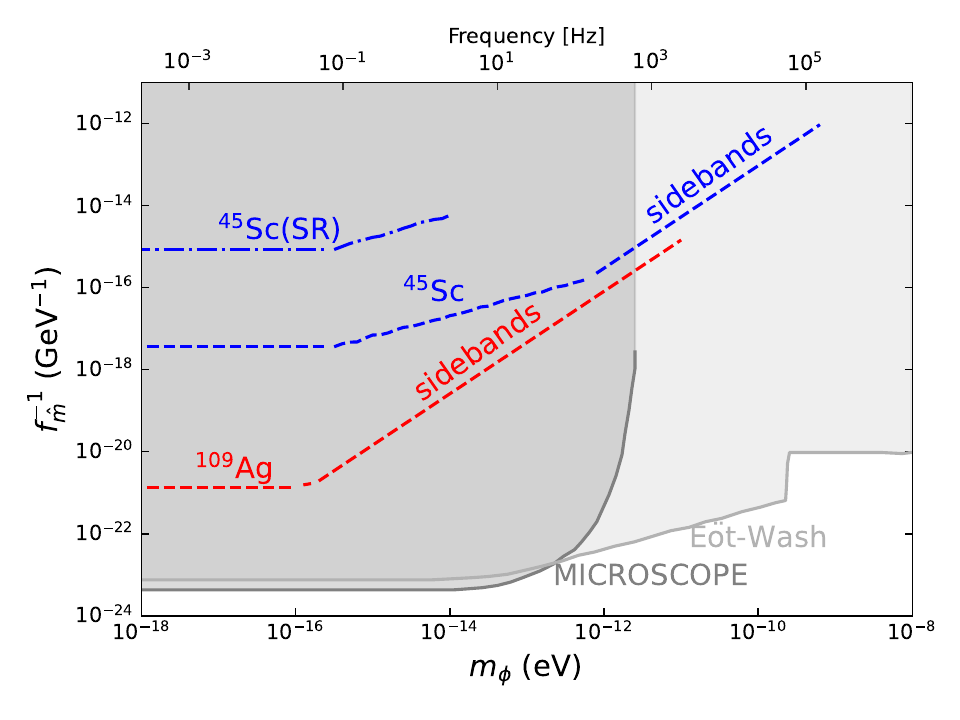}
    \subcaption{(b)}
    \label{fig:f_quark}
\end{subfigure}
\caption{Projected sensitivity to the scalar DM--gluon coupling $f_{g}^{-1}$ (a) and the scalar DM--quarks coupling $f_{\hat{m}}^{-1}$ (b), versus the dark matter mass $m_{\phi}$. 
Red and blue dashed lines denote results for $^{109}\text{Ag}$ and $^{45}\text{Sc}$, respectively. while the blue dot-dashed line corresponds to $^{45}\text{Sc}$ with synchrotron radiation. Approximate sideband sensitivities are shown for comparison. Existing constraints from equivalence-principle tests by MICROSCOPE~\cite{Touboul:2017grn,Berge:2017ovy} and Eöt-Wash~\cite{Smith:1999cr,Schlamminger:2007ht} are shown for comparison. }
\label{Fig:f_gluon_quark}
\end{figure}

We investigate the parameter space for each interaction channel by equating the dark matter–induced energy shifts, given in Eqs.~\ref{eq:delta_E_gamma},~\ref{eq:delta_E_quark}, and \ref{eq:delta_E_gluon}, to the experimentally accessible shift described in Eq.~\ref{eq:delta_h_delta_E}.
An emitter source with an activity of $A= 0.1~\mathrm{Ci}~(3.7~\mathrm{GBq})$ is located at the center, while the absorber and detectors are uniformly distributed along a circle of radius $L = 1~m$ for $^{109}\text{Ag}$ and $^{45}\text{Sc}$. Under these conditions, we derive the corresponding constraints on the DM–photon coupling constant $f_\gamma$ are presented in Fig.\ref{Fig:f_gamma}, the DM--gluon coupling constant $f_g$ are presented in Fig.\ref{fig:f_gluon}, and the DM--quark coupling constant $f_{\hat{m}}^{-1}$ are presented in Fig.\ref{fig:f_quark}, at the $95\%$ confidence level. The sensitivities shown here are based on statistical uncertainties only and do not include detector or systematic effects.
In our analysis, we adopt the parameters listed in Table~\ref{tab:isotope} and Table~\ref{tab:isotope_parameters}. For the traditional M\"ossbauer sources based on $^{109}\text{Ag}$ and $^{45}\text{Sc}$, the experimentally measured linewidths are used in the transmission function given in Eq.~\ref{eq:parametrization}.
For the M\"ossbauer resonance of $^{45}\text{Sc}$ excited by synchrotron radiation, the nuclear resonance photon flux $N_{\gamma}^{\mathrm{MS}}$ in Eq.~\ref{eq:N_gamma_MS_SR} is calculated using the natural linewidth, while the experimentally measured linewidth is adopted in the transmission function Eq.~\ref{eq:parametrization}.

Fig.~\ref{Fig:f_gamma} shows the constraints on the scalar DM--photon coupling $f_{\gamma}^{-1}$. The red and blue dashed lines correspond to $^{109}\mathrm{Ag}$ and $^{45}\mathrm{Sc}$ using radioactive sources, respectively, while the blue dot-dashed line represents the constraint from $^{45}\mathrm{Sc}$ using synchrotron radiation.
For constraints obtained with Mössbauer radioactive sources, the dark matter mass cannot be arbitrarily small, since the effective experimental integration time is limited by the lifetime of the parent isotope. This introduces a low-mass cutoff in the parameter space. Specifically, the dark matter oscillation period $2\pi /m_{\phi}$ must not exceed the lifetime of the parent nucleus $t_P$.
For the $^{109}\mathrm{Ag}$ system, the parent isotope is $^{109}\mathrm{Cd}$ with lifetime $t_{\mathrm{Cd}}$, giving
\begin{equation}
 m_{\phi} > 2\pi /t_{\mathrm{Cd}} \simeq 1.04 \times 10^{-22} \mathrm{eV}.
\end{equation}
Similarly, for the $^{45}\mathrm{Sc}$ system, the corresponding lower bound is $m_{\phi} >  2.94 \times 10^{-22}$eV. 
At high dark matter masses, the sensitivity is limited by the requirement of obtaining sufficient photon counts in each height bin at the $3\sigma$ level.
For the SR scenario, the sensitivity is not constrained by the decay lifetime of a parent isotope, but rather by the realistic operation time available at the synchrotron facility. The ULDM field remains phase coherent only over one coherence time, 
\begin{equation}
\tau \simeq 10^{6}\times \frac{2\pi}{m_{\phi}} \, .
\end{equation}
For $m_{\phi} < 10^6 \times 2\pi/t_{\mathrm{P}}$ and $m_{\phi} \ge 10^6 \times 2\pi/t_{\mathrm{P}}$, we assume an effective experimental integration time, $t_{\rm obs} = t_\mathrm{P}$ and $t_{\rm obs} = \tau $, respectively.
In this work, We conservatively take the observation time to be one dark matter coherence time at high dark matter masses.
Longer observation times, $t_{\rm obs}\gg \tau$, can provide multiple statistically independent measurements and further improve the sensitivity.

In Fig.~\ref{Fig:f_gamma}, the red and blue dashed lines denote the projected sensitivities obtained with $^{109}\mathrm{Ag}$ and $^{45}\mathrm{Sc}$ radioactive sources, respectively, while the blue dot-dashed line corresponds to the result for $^{45}\mathrm{Sc}$ using synchrotron radiation. 
For $^{109}\mathrm{Ag}$, the projected sensitivity can reach approximately $10^{-19}~\mathrm{GeV^{-1}}$. For $^{45}\mathrm{Sc}$, the projected sensitivity is approximately $10^{-15}~\mathrm{GeV^{-1}}$ with a conventional radioactive source and about $10^{-13}~\mathrm{GeV^{-1}}$ in the SR-based configuration.
The green shaded region indicates the GEO 600 bound~\cite{Vermeulen:2021epa}, obtained through spectral analysis of the strain data from the GEO 600 gravitational-wave interferometer. The magenta region shows the exclusion limit from Dynamical Decoupling (DD)~\cite{Aharony:2019iad}, derived from the non-observation of variations in the fine-structure constant $\alpha$ caused by oscillating scalar dark matter in an atomic optical transition. 
The cyan region represents the `DArk Matter from Non Equal Delays' experiment bound~\cite{Savalle:2020vgz}, obtained from searches for dark matter using an optical cavity with unequal delay interferometry. The brown region shows the Holometer bound~\cite{Aiello:2021wlp}, derived by studying variations of the fine-structure constant $\alpha$ using cross-correlated data from the Fermilab Holometer instrument.
The orange region represents constraints derived from $\mathrm{Yb^{+}}$ and Sr atomic clock experiments~\cite{Sherrill:2023zah,Filzinger:2023zrs,Arakawa:2026mls}.
The solid gray and light gray regions correspond to fifth-force (FF) and equivalence-principle (EP) tests, including the MICROSCOPE experiment~\cite{Touboul:2017grn,Berge:2017ovy} and the Cu/Pb torsion-pendulum experiment performed by the Eöt-Wash group~\cite{Smith:1999cr,Schlamminger:2007ht}, following Refs.~\cite{Banks:2024sli,Hees:2018fpg}. Among the nuclei considered here, $^{109}\mathrm{Ag}$ provides the strongest sensitivity. In the high-mass region, its projected sensitivity surpasses the constraints from Dynamical Decoupling (DD), while in the low-mass region it approaches the current limits from EP tests, although it remains weaker than the bounds derived from $\mathrm{Yb^{+}}$ and Sr atomic clock experiments.

%The solid gray and orange lines represent constraints from “fifth-force” (FF) searches and tests of the equivalence principle (EP). The most stringent constraints for this dark matter scenario come from the MICROSCOPE experiment~\cite{Touboul:2017grn,Berge:2017ovy}, indicated by the solid gray lines, and the Cu/Pb torsion pendulum experiment performed by the Eöt-Wash group~\cite{Smith:1999cr,Schlamminger:2007ht}, indicated by the solid orange line, as computed in Refs.~\cite{Banks:2024sli,Hees:2018fpg}.
%Among the considered nuclei, $^{109}\text{Ag}$ exhibits the strongest sensitivity, while the constraint obtained from $^{45}\text{Sc}$ is comparatively weaker.  In the high mass range, the constraint derived from $^{109}\text{Ag}$ reaches a sensitivity level comparable to that achieved by the GEO~600 experiment. In the low mass range, the $^{109}\text{Ag}$ result surpass the limits from equivalence-principle (EP) violation tests, but weaker than the sensitivities of $\mathrm{Yb^{+}}$ and Sr Atomic Clocks. 

As noted in Ref.~\cite{Banerjee:2024bkp}, the homogeneous-shift approximation is valid when the dark-matter oscillation period is longer than the nuclear lifetime, corresponding to the regime $m_{\phi}\lesssim \Gamma$. In this limit, the DM-induced energy shift can be treated as quasi-static and approximately uniform across the resonance line. For both $^{109}\mathrm{Ag}$ and $^{45}\mathrm{Sc}$, the sensitivities presented in this work are valid in this regime.
For $m_{\phi}\gtrsim \Gamma$, however, the rapid oscillation of the dark matter field leads to the appearance of sidebands at $E_{0}\pm n m_{\phi}$ ($n=\pm1,\pm2,\dots$), rather than a simple homogeneous shift of the central resonance line.
The modulation may manifest as an effective symmetric broadening or a small residual shift of the central resonance line. Consequently, the sensitivities would be reduced. As discussed in Ref.~\cite{Fuchs:2024xvc}, the first sidebands located at $E_0\pm m_\phi$ have an intensity suppressed by a factor $\sim (\Delta E/2 m_\phi)^2$, and the intensities of higher order are further suppressed. If the sidebands become detectable, the shift energy $\Delta E$ could be constrained with a scaling as $\propto m_{\phi}/N_{total}^{1/4}$.   
The corresponding sideband sensitivities in Figs.~\ref{Fig:f_gamma} and \ref{Fig:f_gluon_quark} should be regarded as order-of-magnitude estimates. In a dedicated analysis, one would combine the full dataset to reconstruct the complete absorption spectrum and search for the sideband structure around the central resonance frequency. Moreover, at higher dark matter masses, longer observation times, $t_{\rm obs}\gg \tau$, would provide multiple statistically independent measurements and further improve the sensitivity.

%{\color{blue} In this regime, one should combine the full dataset to construct the complete spectrum and search for the sideband structure around the central resonance frequency, rather than dividing the data into segments of duration $\Delta t$. Following the discussion in Ref.~\cite{Fuchs:2024xvc}, we use this scaling to provide rough estimates of the sideband sensitivities shown in Figs.~\ref{Fig:f_gamma} and~\ref{Fig:f_gluon_quark}.}

Similarly, the projected sensitivity can be estimated for scalar DM couplings to gluons and quarks. In each case, only one type of coupling is assumed to be present, while possible interference effects between different couplings are neglected.
%Fig.~\ref{fig:f_gluon} shows the projected constraints on the scalar DM--gluon coupling $f_{g}^{-1}$. The strongest reach is obtained with $^{109}\mathrm{Ag}$, achieving sensitivities down to $10^{-22}~\mathrm{GeV^{-1}}$. The constraints from $^{45}\mathrm{Sc}$ are less stringent. Fig.~\ref{fig:f_quark} presents the projected constraints on the scalar DM--quark coupling $f_{\hat{m}}^{-1}$. Similarly, $^{109}\mathrm{Ag}$ yields the strongest sensitivity, achieving sensitivities down to $10^{-21}~\mathrm{GeV^{-1}}$, while the constraints from $^{45}\mathrm{Sc}$ remain weaker.
Fig.~\ref{fig:f_gluon} shows the projected constraints on the the scalar DM--gluon coupling $f_{g}^{-1}$. The strongest reach is obtained with $^{109}\mathrm{Ag}$, whose best projected sensitivity reaches approximately $10^{-22}~\mathrm{GeV^{-1}}$. The constraints from $^{45}\mathrm{Sc}$ are less stringent.
Fig.~\ref{fig:f_quark} presents the projected constraints on the scalar DM--quark coupling $f_{\hat{m}}^{-1}$. Similarly, $^{109}\mathrm{Ag}$ yields the strongest sensitivity, with the best projected sensitivity reaching approximately $10^{-21}~\mathrm{GeV^{-1}}$, while the constraints from $^{45}\mathrm{Sc}$ remain weaker.

Among the nuclei considered, $^{109}\text{Ag}$ and $^{45}\text{Sc}$, $^{109}\text{Ag}$ yields the strongest constraints on the dark matter couplings, making it the most promising candidate for future precision experiments aimed at probing interactions between dark matter and the Standard Model.
$^{45}\mathrm{Sc}$ with a traditional radioactive source provides comparatively weaker limits. Although the current constraints derived from $^{45}\mathrm{Sc}$ using a synchrotron-radiation-based source are weaker than those obtained from $^{109}\text{Ag}$, this approach offers several important advantages. In particular, the experimental integration time can be extended, and the photon flux may be significantly improved with future advances in synchrotron technology. 
An X-ray free-electron laser oscillator represents a promising next-generation hard X-ray source, capable of delivering fully coherent pulses with meV-level bandwidth and a highly stable flux that is expected to reach $1\times10^{15}\,\mathrm{ph\,s^{-1}\,meV^{-1}}$~\cite{Adams:2019njp}. If applied to dark matter searches, such improvements would substantially increase the number of available photons and could further enhance the experimental sensitivity.
Overall, $^{109}\text{Ag}$ provides the most stringent and robust constraints among the nuclei considered, highlighting its strong potential for future high-precision probes of dark-matter interactions with the Standard Model.

\section{Conclusions and discussion}
\label{sec:conclusion}

The Mössbauer effect is a powerful tool for detecting small shifts in nuclear energy levels. In this study, we explore its potential for probing the coupling between dark matter and nuclei in the source and absorber of a Mössbauer spectrometer. To investigate possible variations in nuclear transition energies induced by an oscillating dark matter background, we consider a static M\"ossbauer  measurement scheme to probe nuclear transition energy variations induced by an oscillating scalar dark matter field. The ultralight dark matter interacting with both the emitter and absorber can lead to a variation in their nuclear energy levels, resulting in a measurable shift in the position of the Mössbauer resonance.

In our analysis, we select $^{109}\text{Ag}$ and $^{45}\text{Sc}$ as candidate Mössbauer isotopes. We assume an emitter source with an activity of $A= 0.1~\mathrm{Ci}\,(3.7~\mathrm{GBq})$, and also consider $^{45}\text{Sc}$ with synchrotron radiation scheme.
The baseline distance between  emitter and absorber for $^{109}\text{Ag}$ and $^{45}\text{Sc}$ is $L=1~m$.
The DM mass is varied in the range from $10^{-20}\mathrm{eV}$ to $10^{-8}~\mathrm{eV}$. Under these assumptions, we derive projected constraints on the DM--photon coupling $f_\gamma^{-1}$, the DM--gluon coupling $f_g^{-1}$, and the DM--quark coupling $f_{\hat{m}}^{-1}$.

Among the isotopes considered in this work, $^{109}\mathrm{Ag}$ provides the strongest sensitivity, while $^{45}\mathrm{Sc}$ yields comparatively weaker constraints, with the synchrotron-radiation-based $^{45}\mathrm{Sc}$ setup giving the weakest reach.
For $^{109}\mathrm{Ag}$, the projected sensitivity for the scalar DM--photon coupling $f_{\gamma}^{-1}$ reaches approximately $10^{-19}~\mathrm{GeV^{-1}}$, approaching the current limits from EP tests. For the scalar DM--gluon coupling $f_{g}^{-1}$ and the scalar DM--quark coupling $f_{\hat{m}}^{-1}$, the projected sensitivities of $^{109}\mathrm{Ag}$ reach approximately $10^{-22}$ and $10^{-21}~\mathrm{GeV^{-1}}$, respectively.
Although the physically viable parameter space accessible in the present study is already covered by existing experimental constraints from equivalence principle tests, searches based on the M\"ossbauer effect provide a complementary experimental avenue for probing ultralight dark matter.  
Owing to their distinct detection principle and ultra-high energy resolution, such experiments offer an independent test of scalar dark matter interactions and broaden the range of experimental techniques available for exploring physics beyond the Standard Model.

The $^{67}\mathrm{Zn}$ isotope can also be considered as an alternative candidate. Its $93.3~\mathrm{keV}$ transition has a relatively narrow experimental linewidth of $7.5 \times 10^{-11}~\mathrm{eV}$~\cite{Gratta:2020hyp}, which leads to a gravitational broadening of $\delta Z = 7.4~\mathrm{m}$. Accordingly, the detector spatial resolution would need to be set to $3.7~\mathrm{m}$. This would require a longer baseline and a larger detector, and photon number attenuation along the long propagation distance must also be taken into account. While this transition could in principle provide meaningful constraints, its experimental feasibility remains challenging. Future studies could explore the potential of $^{67}\mathrm{Zn}$ for such measurements.

%In this work, we have considered a table-top baseline distance of $L = 1~\mathrm{m}$ for $^{109}\text{Ag}$. In the long DM wavelength limit, extending the baseline to 10~m or 100~m linearly increases the energy shift at the cost of lowering statistics at the detectors. As the significance depends on photon counting as $\sim 1/\sqrt{N}$, a longer baseline still expects an improvement to the sensitivity.

Further enhancements could be achieved by increasing the source activity and incorporating optic control, both of which would increase the number of detected photons and thereby improve the photon statistics.
In light of future powerful synchrotron radiation sources  delivering high-brightness X-ray beams, synchrotron-based M\"ossbauer spectroscopy has become increasingly important. In particular, the development of narrow-band, high-repetition-rate XFEL has enabled resonant excitation of $^{45}\text{Sc}$, as recently demonstrated at the European XFEL. If such technology were applied to dark matter searches, it could substantially increase the number of emitted photons and further enhance the experimental sensitivity.

\section*{Acknowledgments}
The authors thank Hua-Qiao Zhang, Xing-Jian Lv and Han Miao for valuable discussions.
This work is supported by the National Natural Science Foundation of China (Grants No. 12175248, 12447105 and 12275067), 
Science and Technology R$\&$D Program Joint Fund Project of Henan Province  (Grant No.225200810030),
Science and Technology Innovation Leading Talent Support Program of Henan Province, and National Key R$\&$D Program of China (Grants No. 2023YFA1606000 and 2023YFA1608903).
During the preparation of this paper, ~\cite{liu2025probingaxionmossbauerspectroscopy} appeared with a focus on the pseudo-scalar type of ultra-light dark matter.

%.W.X was supported by the National Key R&D Program of China (Grants  No. 2023YFA1608903）

\bibliographystyle{apsrev4-2}
\bibliography{bibliography}

\end{document}